\documentclass[prd,twocolumn,showpacs,amsmath,amssymb,floatfix]{revtex4}
\usepackage{graphicx,color,dcolumn}
\begin{document}
\title{The line shape of the radiative open-charm decay of $Y(4140)$ and $Y(3930)$}

\author{Xiang Liu\footnote{Corresponding author}}\email{xiangliu@lzu.edu.cn}
\affiliation{School of Physical Science and Technology, Lanzhou University, Lanzhou 730000,  China}
\author{Hong-Wei Ke$^{*}$}\email{khw020056@hotmail.com}
\affiliation{Physical Department, School of Science, Tianjin University, Tianjin 300072, China}

\date{\today}
\begin{abstract}
In this work, we study the radiative open-charm decays $Y(4140)\to {D}_s^{\ast+} D_s^- \gamma$ and $Y(3930)\to{D}^{\ast+} D^-\gamma$
under the assignments of $D_{s}^*\bar{D}_s^*$ and $D^*\bar{D}^*$ as molecular states for $Y(4140)$ and $Y(3930)$ respectively. Based on our numerical result, we propose the experimental measurement of the photon spectrum of $Y(4140)\to { D}_s^{\ast+} D_s^- \gamma,\,D_{s}^+D_{s}^{*-}\gamma$ and $Y(3930)\to
D^{*0}\bar{D}^0\gamma,\,D^{0}\bar{D}^{*0}\gamma,\,
D^{*+}D^-\gamma,\,D^+D^{*-}\gamma$ can further test the molecular assignment for $Y(4140)$ and $Y(3930)$.
\end{abstract}

 \maketitle


In the past six years, experiments have announced a series of
charmonium-like states $X(3872)$, $X(3930)/Y(3930)/Z(3930)$,
$Y(4260)$, $Z^+(4430)$ etc.. Recently the CDF Collaboration
reported the observation of $Y(4140)$ by studying the invariant
mass spectrum of $J/\psi \phi$ in $B^+\to K^+ J/\psi \phi$. Its
mass and width are
$m=4143.0\pm2.9(\mathrm{stat}.)\pm1.2(\mathrm{syst}.)$ MeV and
$\Gamma=11.7^{+8.3}_{-5.0}(\mathrm{stat}.)\pm3.7(\mathrm{syst}.)$
MeV, respectively \cite{Aaltonen:2009tz}. $Y(4140)$ not only makes
the spectroscopy of the charmonium-like state abundant, but also
provides a good chance to further understand the property of the
observed charmonium-like states.

In the observed charmonium-like states, $Y(3930)$ is a near-threshold
$\omega J/\psi$ mass enhancement in the exclusive $B\to K \omega
J/\psi$ decays, which was firstly observed by the Belle
Collaboration \cite{Abe:2004zs} and confirmed by the Babar
Collaboration \cite{Aubert:2007vj}. Since both $Y(4140)$ and
$Y(3930)$ were observed in the mass spectrum of $J/\psi+V$ of
$B\to K J/\psi V$ channel ($V$ denotes light vector meson),
$Y(4140)$ is similar to $Y(3930)$.

The authors of Ref. \cite{Liu:2009ei} discussed the various
possible interpretations of the $Y(4140)$, and further proposed
that $Y(4140)$ is probably a $D^*_s\bar{D}_s^*$ molecular state
with $J^{PC}=0^{++}$ or $J^{PC}=2^{++}$ while $Y(3930)$ is its
$D^*\bar{D}^*$ molecular partner. Relevant dynamics calculation of
$D^*_s\bar{D}_s^*$ and $D^*\bar{D}^*$ was performed
in the potential model \cite{Liu:2008tn}.
Later N. Mahajan argued $Y(4140)$ to be a $D^*_s\bar{D}_s^*$
molecular state or an exotic ($J^{PC} = 1^{-+}$) hybrid charmonium
\cite{Mahajan:2009pj}.

Besides using the potential model to dynamically study
$D^*_s\bar{D}_s^*$ system, QCD sum rule (QSR) is applied to
calculate the mass spectrum of $D^*_s\bar{D}_s^*$ system
\cite{Wang:2009ue,Albuquerque:2009ak,Zhang:2009st}. The mass of
$D_{s}^*\bar{D}_s^*$ system from the QSR calculation in Ref.
\cite{Wang:2009ue} is not consistent with the experimental value
of $Y(4140)$. The authors in Ref.
\cite{Albuquerque:2009ak,Zhang:2009st} also used QSR to calculate
the mass of $D_{s}^*\bar{D}_{s}^*$ system. Their numerical results
indicate the existence of a $D_{s}^*\bar{D}_{s}^*$ bound state,
which is consistent with CDF experimental observation and does not
support the result in Ref. \cite{Wang:2009ue}.

By using one boson exchange model, Ding obtained the effective
potential of $D_s^*\bar{D}_s^*$ system \cite{Ding:2009vd}. The
result supports to explain $Y(4140)$ as a $D_{s}^*\bar{D}_s^*$
molecular state with the quantum number $J^{PC}=0^{++}$. Meanwhile,
searching the $1^{+-}$ and $1^{--}$ partners of $Y(4140)$ in
$J/\psi \eta$ and $J/\psi\eta^\prime$ were suggested
\cite{Ding:2009vd}.

As indicated in Ref. \cite{Liu:2009ei}, tetraquark $Y(4140)$ falls apart into a pair of charmed mesons
very easily under the assignment of tetraquark. In general, the width of the tetraquark would be broad,
which does not consist with the experimental value $\Gamma=12$ MeV. Recently Stancu calculated the spectrum of $c\bar{c}s\bar{s}$ by a quark model with chromomagnetic interaction \cite{Stancu:2009ka}, by considering a tetraquarks assignment ($c\bar{c}s\bar{s}$) for $Y(4140)$, which favors $J^{PC}=1^{++}$. In this assignment, the coupling constant of $Y(4140)$ with $VV$ channel is small, which can alleviate the contradiction
between the small experimental width and the large width resulted from the fall apart mechanism.

There exist different understandings
from the exotic explanation to the source of $Y(4140)$. In recent work \cite{vanBeveren:2009dc},
van Beveren and Rupp proposed that the $Y(4140)$ enhancement resulted from the opening of the $J/\psi\phi$ channel and that probably does not represent a resonance.

As an important and interesting topic, studying the decay of
$Y(4140)$ can be helpful to reveal its underlying structure. By using an
effective Lagrangian approach, Branz, Gutsche and Lyubovitskij
\cite{Branz:2009yt} calculated $Y(3930)\to J/\psi\omega$ and
$Y(4140)\to J/\psi \phi$ strong decays and $Y(3930)/Y(3940)\to
\gamma\gamma$ decays, which are induced by the hadronic loop effect.
Furthermore, the molecular explanation for $Y(3930)$ and $Y(4140)$
is supported by the result of $J/\psi V$ ($V=\omega,\phi$) mode of
$Y(3930)$ and $Y(4140)$ \cite{Branz:2009yt}.

In Ref. \cite{Liu:2009iw}, one of the authors of this work calculated
the hidden charm decay of $Y(4140)$ in the assumption of the
second radial excitation of the P-wave charmonium
$\chi_{cJ}^{\prime\prime}$ ($J =0,\,1$). Since the branching ratio
of the hidden charm decay $Y(4140)\to J/\psi\phi$ is of the order
of $10^{-4}\sim 10^{-3}$ under the $\chi_{cJ}^{\prime\prime}$ assignment for $Y(4140)$, which disfavors the large hidden charm
decay pattern indicated by the CDF experiment, the pure second radial excitation of the P-wave charmonium
$\chi_{cJ}^{\prime\prime}$ ($J=0,\,1$) is problematic
\cite{Liu:2009iw}.

Besides investigating the hidden charm decay, the open charm decay and
the double gamma decay of $Y(4140)/Y(3930)$, the radiative decay of
$Y(4140)/Y(3930)$ can provide the useful information to distinguish the
molecular explanation of $Y(4140)/Y(3930)$ from other assignments,
which was indicated in Ref. \cite{Liu:2009ei}. The typical
radiative decay modes are ${\bar D}^\ast D\gamma$, ${\bar D}\gamma
D\gamma$, ${\bar D}\pi D\gamma$ for $Y(3930)$, and ${\bar
D}_s^\ast D_s \gamma$, ${\bar D}_s\gamma D_s \pi^0$, ${\bar
D}_s\gamma D_s \gamma$ for $Y(4140)$.

Due to the suppressions from the phase space and the $\alpha=1/137$ of the electromagnetic vertex, the width
of the four-body radiative decay of $Y(4140)/Y(3930)$ is far less
than that of the three-body radiative decay of $Y(4140)/Y(3930)$.
Thus, in this work we will be dedicated to the study of the three-body
radiative decays $Y(3930)\to{\bar D}^\ast D\gamma$ and $Y(4140)\to
{\bar D}_s^\ast D_s \gamma$, which are named as the radiative
open-charm decay. The radiative open-charm decays of $Y(4140)$ are
$D_s^{*+}D_s^-\gamma$ and $D_{s}^+D_{s}^{*-}\gamma$. For
$Y(3930)$, its radiative open-charm decays include
$D^{*0}\bar{D}^0\gamma$, $D^{0}\bar{D}^{*0}\gamma$,
$D^{*+}D^-\gamma$ and $D^+D^{*-}\gamma$. In the following, we will
illustrate the radiative open-charm decays of $Y(4140)$ and
$Y(3930)$ with $Y(4140)\to D_s^{*+}D_s^-\gamma$ and $Y(3930)\to
D^{*+}D^-\gamma$ as the example.

\begin{center}
\begin{figure}[htb]
\begin{tabular}{cccccccc}
\scalebox{0.45}{\includegraphics{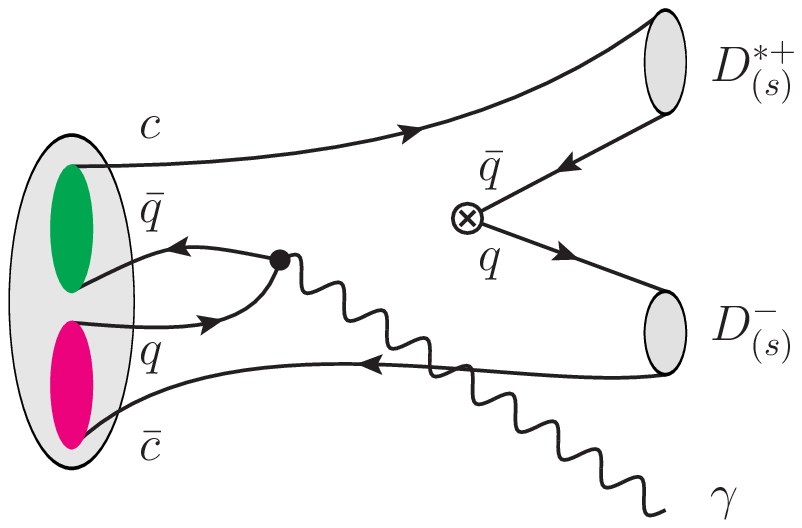}}&&&&\raisebox{0.6em}{\scalebox{0.45}{\includegraphics{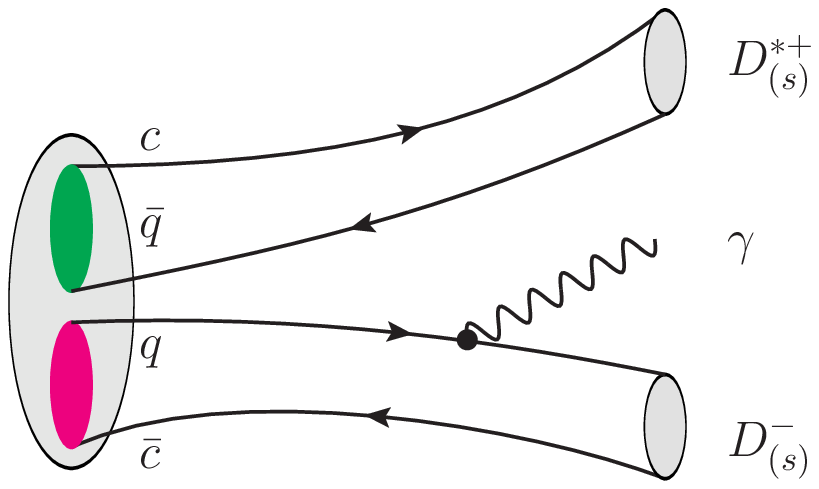}}}
\\ \\
(a)&&&&(b)\\ \\
\raisebox{0.8em}{\scalebox{0.45}{\includegraphics{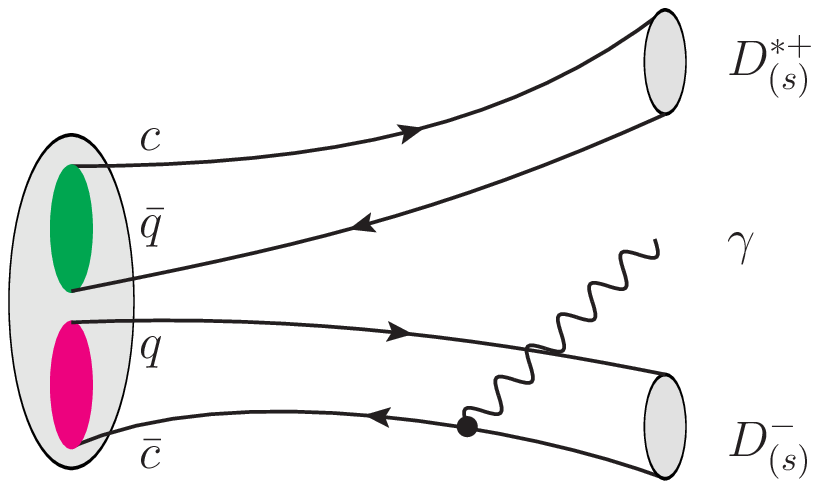}}}\\(c)&&
\end{tabular}
\caption{The diagrams depicting the radiative open-charm decay of $Y(4140)$ and $Y(3930)$ in the quark level. \label{quark-diagram}}
\end{figure}
\end{center}

The radiative open-charm decay of $Y(4140)$ and $Y(3930)$ occurs via the two different mechanisms, which are depicted by Fig. \ref{quark-diagram} in the quark level:

\noindent $\bullet$ {\bf Mechanism A} The $q$ and $\bar{q}$ respectively in the molecular components $D_{(s)}^{*-}$ and $D_{(s)}^{*+}$ annihilate into a photon $\gamma$. Then the rest $c$ and $\bar{c}$ respectively combine with $\bar q$ and ${q}$ created from vacuum by Quark Pair Creation (QPC) mechanism \cite{Micu:1968mk,Le Yaouanc:1972ae} to form $D_{s}^{*+}$ and $D_{s}^{-}$, which is depicted by Fig. \ref{quark-diagram} (a).

\noindent $\bullet$ {\bf Mechanism B} By emitting a photon $\gamma$ from $q$ or $\bar{c}$, the molecular state collapses into $D_{s}^{*+}$ and $D_{s}^{-}$, which are described in Fig. \ref{quark-diagram} (b) and (c). In fact, the decay mechanism depicted in Fig. \ref{quark-diagram} (b) and (c) could be as well used in a tetraquark interpretation.
Due to the difference of the wave functions describing $Y(4140)$ under the molecular structure assignment and the tetraquark assignment \cite{Liu:2009ei,Stancu:2009ka}, we can distinguish the radiative decay under molecular state and tetraquark assignments. For the molecular state explanation, we will use an S-wave wave function to represent the interaction between $D_{(s)}^{*-}$ and $D_{(s)}^{*+}$. In the following, we will illustrate it more explicitly.

Since there exists an extra quark pair creation for the process resulted from Mechanism A, the rate of the process depicted by Mechanism A is relatively suppressed comparing with that described by Mechanism B \cite{Micu:1968mk,Le Yaouanc:1972ae}. Thus $Y(4140)\to D_s^{*+}D_s^-\gamma$ and $Y(3930)\to D^{*+}D^-\gamma$ are dominantly controlled through Mechanism B.

\begin{figure}[htb]
\scalebox{0.5}{\includegraphics{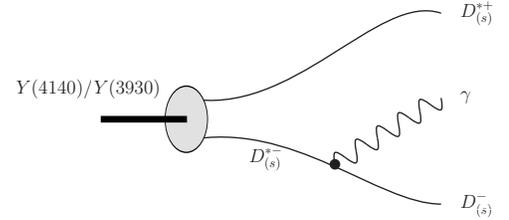}}
\caption{The hadron level description for $Y(4140)\to D_s^{*+}D_s^-\gamma$ and $Y(3930)\to D^{*+}D^-\gamma$
under the molecular assumption. \label{fig:radiative}}
\end{figure}

An alternative description in the hadron level for $Y(4140)\to D_s^{*+}D_s^-\gamma$ and $Y(3930)\to D^{*+}D^-\gamma$ depicted by Mechanism B is presented by Fig. \ref{fig:radiative}. In the hadron level, $Y(4140)\to D_s^{*+}D_s^-\gamma$ and $Y(3930)\to D^{*+}D^-\gamma$ occur via the radiative decays of the intermediate states $D_{s}^{*-}$ and $D^{*-}$ respectively.
The general transition matrix element for $Y(3930)\to{D}^{\ast+} D^-\gamma$ and
$Y(4140)\to { D}_s^{\ast+} D_s^- \gamma$ processes can be expressed as
\begin{eqnarray}
&&\mathcal{M}[Y(3930)(Y(4140))\to D_{(s)}^{*+}D_{(s)}^-\gamma]\nonumber\\&&=\langle D_{(s)}^-\gamma |\mathcal{H}_2|D_{(s)}^{*-}\rangle
 \langle D_{(s)}^{*+}D_{(s)}^{*-} |\mathcal{H}_1|Y\rangle,
\end{eqnarray}
where $\mathcal{H}_1$ describes the collapse of S-wave $D_{(s)}^{*+}{D}_{(s)}^{*-}$ molecular state.
$\mathcal{H}_2$ denotes the interaction of $D_{(s)}^{*-}$ and $D_{(s)}^-\gamma$.
For describing the decay amplitude, we adopt the same method as that in Refs. \cite{Voloshin:2003nt,Voloshin:2005rt}
to study the radiative open-charm decays of $Y(4140)$ and $Y(3930)$. In Refs. \cite{Voloshin:2003nt,Voloshin:2005rt}, Voloshin once assumed $X(3872)$ is a $D\bar{D}^*+h.c.$ molecular
state and explored the radiative decay $X(3872)\rightarrow
D\bar{D}\gamma$.

The matrix element $\langle D_{(s)}^{*+}D_{(s)}^{*-} |\mathcal{H}_1|Y\rangle$, describing the collapse of $Y(4140)$ into $D_{(s)}^{*+}$ and $D_{(s)}^{*-}$, can be represented as
\begin{eqnarray}
\langle D_{(s)}^{*+}D_{(s)}^{*-} |\mathcal{H}_1|Y\rangle=\alpha \mathcal{C}\Psi(\vec q),
\end{eqnarray}
where $\alpha$ is the weight of the corresponding component
$|D_{(s)}^{*+}D_{(s)}^{*-}\rangle$ in the molecular wave function
of $Y(4140)/Y(3930)$  \cite{Liu:2009ei}
\begin{eqnarray}
|Y(4140)\rangle&=&|D_s^{*+}D_s^{*-}\rangle,\label{111}\\
|Y(3930)\rangle&=&\frac{1}{\sqrt{2}}\Big[|D^{*0}\bar{D}^{*0}\rangle+|D^{*+}D^{*-}\rangle\Big],\label{222}
\end{eqnarray}
respectively. $\Psi(\vec{q})$ is the
wave function describing S-wave $D_{(s)}^{*+}D_{(s)}^{*-}$ molecular state, which is of the form \cite{Voloshin:2003nt,Voloshin:2005rt}
\begin{eqnarray}
\Psi(\vec{q})=\sqrt{8\pi\kappa}\Big(\frac{1}{q^2+\kappa^2}\Big)
\end{eqnarray}
with $\kappa=\sqrt{2m_\tau E}$. Here $E$ and $m_\tau$ are the binding energy
and the reduced mass of $D_{(s)}^{*+}D_{(s)}^{*-}$ molecular system respectively.
$\vec{q}$ is the relative momentum between $D_{(s)}^{*-}$
and $D_{(s)}^{*+}$ in the molecular system. Factor $\mathcal{C}$ represents the polarization effect from
$Y(3930)(Y(4140))$ and $D_{(s)}^{*\pm}$.

The amplitude describing $D_{(s)}^*\rightarrow D_{(s)}\gamma$
reads as
\begin{eqnarray}
\mathcal{A}[D_{(s)}^{*-}\rightarrow
D_{(s)}^-\gamma]=g\,\epsilon_{ij\ell}\varepsilon_{1i}k_j\varepsilon_{2\ell},
\end{eqnarray}
where $g$ is the effective coupling constant. $\vec{k}$ and
$\vec\varepsilon_{1}$ are the three momentum and the polarization
vector of photon, respectively. $\vec\varepsilon_{2}$ denotes the
polarization vector of $D_{(s)}^*$.

Thus the amplitude of decay $Y(4140)\rightarrow D_s^{*+}{D}_s^-\gamma$
or $Y(3930)\rightarrow D^{*+}{D}^-\gamma$ can be written as
\begin{eqnarray}
&&\mathcal{M}[Y(3930)(Y(4140))\rightarrow {D}_{(s)}^{+*}
D_{(s)}^-\gamma]\nonumber\\&&=g\,\alpha\,\epsilon_{ij\ell}
\varepsilon_{1i}k_j\varepsilon_{2\ell}\,\mathcal{C}\,\Psi(\vec{q}) ,
\end{eqnarray}
where $\alpha=1$ and $\alpha=1/\sqrt{2}$ correspond to
$Y(4140)\rightarrow D_s^{*+}{D}_s^-\gamma$ and $Y(3930)\rightarrow
D^{*+}{D}^-\gamma$ processes, respectively.

Finally one obtains the decay rate of $Y(3930)(Y(4140))\rightarrow D_{(s)}^{*+}
{D}_{(s)}^-\gamma$
\begin{eqnarray}
&&d\Gamma[Y(3930)(Y(4140))\rightarrow D_{(s)}^{*+}
{D}_{(s)}^-\gamma]\nonumber\\&&=\frac{2g^2k^2\,\alpha^2\,\mathcal{C}^2}{3(2\pi)^5}\Psi(\vec{q})^2
\delta(E_1+E_2+\omega_0-M)\nonumber\\
&&\quad\times\delta(\vec{p_1}+\vec{p_2}+\vec{k})d^3p_1d^3p_2\frac{d^3k}{(2k)}.
\end{eqnarray}
After integration we obtain the differential decay rate in terms of
the radiative width of $D_{(s)}^{*-}$
\begin{eqnarray}\label{dgdp}
&&d\Gamma[Y(3930)(Y(4140))\rightarrow D_{(s)}^{*+}
{D}_{(s)}^-\gamma]\nonumber\\&&=\frac{g^2\omega^3\alpha^2\,\mathcal{C}^2}{3\pi}\Big[\phi({\vec{k}}+2\vec{p})\Big]^2\frac{d^3p}{(2\pi)^3},
\end{eqnarray}
where $\vec{p}$ is the
momentum of $D_{(s)}^-$ in the c.m. frame of the
$D_{(s)}^-D_{(s)}^{*+}$ system. $\vec{k}$ denotes the momentum of
the photon in the c.m. frame of $Y(4140)/Y(3930)$.
$\vec{q_1}(\vec{q_2}$) and $E_1(E_2)$ are the momentum and
energy of the final meson $D_{(s)}^-$($D_{(s)}^{*+}$). The photon
energy $\omega$ and the momentum $\vec{p}$ are related by
{\small
\begin{eqnarray}
&&|\vec{p}|\nonumber\\&&=\frac{\sqrt{[m_{12}^2-(m_{D_{(s)}^-}+m_{D_{(s)}^{*-}})^2]
[m_{12}^2-(m_{D_{(s)}^-}-m_{D_{(s)}^{*-}})^2]}}{2\sqrt{M^2-2M\omega}}\nonumber\\
\end{eqnarray}}
with the relation $m_{12}=M^2-2M\omega$, where $M$ denotes the mass of inial state $Y(4140)/Y(3930)$.
By using the above relation, we can replace $dp$ with $d\omega$ in
eq. \ref{dgdp}. Further we get the expression of $d\Gamma[Y(3930)(Y(4140))\rightarrow D_{(s)}^{*+}
{D}_{(s)}^-\gamma]/d\omega$. By this expression, we carry out the study of the line shape of the
photon spectrum of $Y(3930)\to{D}^{\ast+} D^-\gamma$ and
$Y(4140)\to { D}_s^{\ast+} D_s^- \gamma$ under the assumption of the $D_s^*\bar{D}_s^{*}/D^{*}\bar{D}^{*}$ molecular state for $Y(4140)/Y(3930)$.

\begin{figure}[htb]
\scalebox{0.8}{\includegraphics{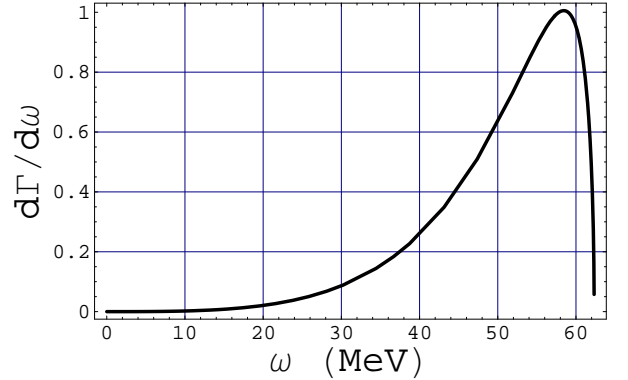}}
\caption{Photon spectrum in $Y(4140)\rightarrow
D_s^{*+}D_s^-\gamma$. Here the maximums of $d\Gamma/d\omega$
is normalized to one. \label{fig:4140}}
\end{figure}

\begin{figure}[htb]
\scalebox{0.8}{\includegraphics{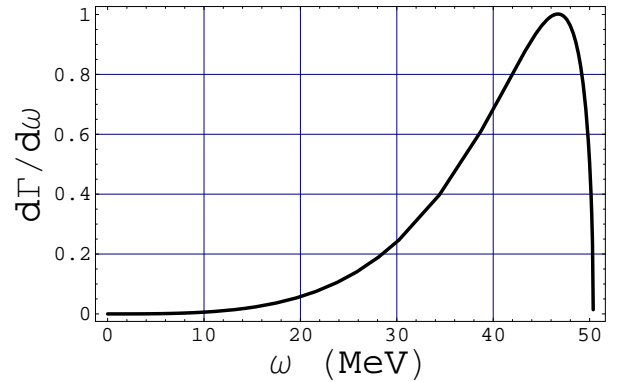}}
\caption{Photon spectrum in $Y(3930)\rightarrow D^{*+}D^-\gamma$. Here the maximums of $d\Gamma/d\omega$
is normalized to one.
\label{fig:3930}}
\end{figure}

In Figs. \ref{fig:4140} and \ref{fig:3930}, the line shape of $d\Gamma[Y(3930)(Y(4140))\rightarrow D_{(s)}^{*+}
{D}_{(s)}^-\gamma]/d\omega$ as a function of the photon energy $\omega$ are presented.
The line shape of the photon spectrum $Y(4140)\to { D}_s^{\ast+} D_s^- \gamma$ is similar to that of $Y(3930)\to{D}^{\ast+} D^-\gamma$.
There exits a sharp peak near the large end-point of photon energy in Figs. \ref{fig:4140} and \ref{fig:3930}.
On the left side of the peak, the photon spectrum changes smoothly with the variation of the photon energy $\omega$ while on the right side of the peak the line shape of the photon spectrum goes down very rapidly with the increasing $\omega$. These features shown in our results can provide useful information for testing the molecular state structure assignment for $Y(4140)$ and $Y(3930)$.

In summary, as indicated in Ref. \cite{Liu:2009ei}, the line shapes of the photon spectrum of $Y(4140)\to { D}_s^{\ast+} D_s^- \gamma$ and $Y(3930)\to{D}^{\ast+} D^-\gamma$ are crucial to test the molecular state assignment for $Y(4140)$ and $Y(3930)$. In this work, we study the photon spectrum of the radiative open-charm decays of $Y(4140)$ and $Y(3930)$ under the structure of the molecular state. Due to the peculiar characters of the line shape of the photon spectrum $Y(4140)\to { D}_s^{\ast+} D_s^- \gamma$ and $Y(3930)\to{D}^{\ast+} D^-\gamma$ shown in our numerical result, we suggest the experimentalist to carry out the measurement of the photon spectrum of the radiative open-charm decay of $Y(4140)$ and $Y(3930)$ in future experiment.

\vspace{1cm}

\noindent {\bf Acknowledgement} We thank
Prof. Shi-Lin Zhu for useful suggestions. We
acknowledge National Natural Science Foundation of China under Grants 10705001 and
A Foundation for the Author of National Excellent Doctoral Dissertation of P.R. China (FANEDD).

\end{document}